\documentclass[referee]{aa} 
\usepackage{graphicx}
\usepackage{txfonts}
\begin{document}
   \title{Solar coronal differential rotation from XBPs in Hinode/XRT and Yohkoh/SXT images}

   \author{R. Kariyappa}

   \institute{Indian Institute of Astrophysics, Bangalore 560 034, India\\
              \email{rkari@iiap.res.in}}
\offprints{R. Kariyappa}

   \date{Received ........................ / Accepted ....................... }

 
  \abstract
{}  
   {Our aim is to identify and trace the X-ray Bright Points
(XBPs) over the disk and use them as tracers to determine the coronal rotation.  This investigation
will help to clarify and understand several issues: whether (i) the corona rotates differentially; 
(ii) the rotation depends on the sizes of the XBPs; and (iii) dependence on phases of the solar magnetic cycle.}
   {We analysed the daily full-disk soft X-ray images observed with (i) X-Ray Telescope
(XRT) on-board the Hinode mission during January, March and April, 2007 and (ii) Soft X-ray Telescope (SXT)
on-board the Yohkoh from 1992 to 2001 using SSW in IDL.  We have used the tracer method to trace the passage of XBPs 
over the solar disk with the help of overlaying grids and derived the sidereal angular 
rotation velocity and the coordinates (latitude and longitude) of the XBPs.}
   { We have determined the position of a large number of XBPs both in Hinode/XRT and Yohkoh/SXT images
and followed them over the solar disk as a function of time.  We derived the coronal sidereal angular
rotation velocity and compared it with heliocentric latitude and as a function of solar activity cycle.  
In addition, we measured
the sizes of all the XBPs and related them with the coronal rotation.  The important results derived from 
these investigations are: (i) the solar corona rotates 
differentially like the photosphere and chromosphere; (ii) the sidereal angular rotation velocity is independent of the sizes
of the XBPs; (iii) the sidereal angular rotation velocity does not depend on phases of 
the solar magnetic cycle; and (iv) the differential rotation of the corona is present
throughout the solar magnetic cycle.} 
   {}


\titlerunning {Solar coronal rotation from XBPs}
\authorrunning {R. Kariyappa}
   \maketitle
    
%

\section{Introduction}

The Sun rotates differentially (i.e. the equatorial regions rotate faster than the polar regions) 
at the photosphere and chromosphere.  The measurements of solar rotation have been carried out by
two methods, namely, (i) Tracer Method - by tracing the passage of various features like sunspots, faculae,
filaments etc., over the solar disk and (ii) Doppler Method - by the spectroscopic observations of Doppler
displacements of the core of the spectral lines.  The phenomenon of solar rotation is still
to be understood clearly from the existing large volumes of data.  On the other hand, the coronal
rotation is observed through features like Fe XIV green lines, soft X-rays and radio waves. 
The coronal rotation has been measured by two methods: (i) Tracer Method - based on visual tracing of 
coronal features (XBPs) in consecutive images and (ii) Automatic Method - relies on the IDL procedure "Regions 
Of Interest" segmentation which is used to identify and follow them in the consecutive images.   However, in
both the methods, there are advantages and disadvantages.
The coronal rotation determination appears to be more complicated and even less understood, because the 
corona is optically thin across a wide range of observed wavelengths and the features are less 
distinct in duration and extent.

Solar coronal X-ray bright points (XBPs) were discovered using a soft X-ray telescope
(SXT) on a sounding rocket and their nature
has been an enigma since their discovery in late 1960's (Vaiana et al. 1970).
Later, using Skylab and Yohkoh X-ray images, the XBPs
were studied in great detail (Golub et al. 1974; Harvey 1996; Nakakubo \& Hara 1999
; Longcope et al. 2001; Hara \& Nakakubo 2003).  It has been shown that the XBPs have a spatial
correspondence with small-scale bipolar magnetic regions by comparing the ground-based magnetic field 
measurements with simultaneous space-born X-ray imaging observations (Krieger et al. 1971; Golub et al. 1977).  
The number of XBPs (daily) found on the Sun varies from several
hundreds upto a few thousands (Golub et al. 1974).  Zhang et al. (2001) found a density of
800 XBPs for the entire solar surface at any given time.  It is known that the observed XBP number
is anti-correlated with the solar cycle, but this is an observational bias and the number
density of XBPs is nearly independent of the 11-yr solar activity cycle (Nakakubo \& Hara 1999;
Sattarov et al. 2002; Hara \& Nakakubo 2003).  Golub et al. (1974) found that the diameters of the
XBPs are around 10-20 arc sec and their life times range from 2 hours to 2 days (Zhang et al. 2001).  The
physical studies have indicated the temperatures to be fairly low, $T= 2 \times 10^6$ K, and
the electron densities
{$n_e$ }= 5 $\times 10^{9} cm^{-3}$ (Golub \& Pasachoff 1997).
Recently, Kariyappa and Varghese 
(2008) have analysed the time sequence of XBPs in soft X-ray images obtained from Hinode/XRT and 
discussed the nature of intensity oscillations associated with the different classes of XBPs showing 
the different emission levels.  It has been shown by various groups that the solar corona rotates 
differentially on the basis of their analysis of coronal bright points in SOHO/EIT images
(Brajsa et al. 2001, 2002, 2004; Karachik et al. 2006), and by using radio observations at different
frequencies (Vats et al. 2001). 
However, further investigations are required to confirm the differential rotation of the 
corona.  Particularly, it is not clear that the sidereal angular
rotation velocity of the corona depends on the sizes and lifetimes of the tracers (XBPs) and how  
it varys with the phases of the solar cycle.
	
In this paper, we have analysed two sets of observations of full-disk soft X-ray images obtained from
Hinode/XRT and Yohkoh/SXT experiments to determine the coronal rotation using the XBPs as tracers and explore the
evidence for differential rotation and its relation to the sizes of the XBPs and phases 
of the solar magnetic cycle.

\section{Observations and methods of data analysis}


\subsection{Hinode/XRT full-disk images}

We have used daily full-disk soft X-ray images obtained during the period of January, March and April, 2007
from X-Ray Telescope (XRT) on-board the Hinode mission.  The images have been observed 
through a single X-ray Ti\_poly filter and the image size is 2048"x2048" with a spatial 
resolution of 1.032"/pixel.  
All the full-disk X-ray images were visually inspected
and the XBPs have been identified in sequences of images.   The main criterion for the identification of a XBP was checking
its persistence in the consecutive images at approximately the same latitude and shifted in the Central Meridian
Distance (CMD) according to the elapsed time.  CMD values of the identified features were then measured in
selected consecutive images and were fitted as a function of time.  The correlation coefficient of the function CMD
was generally very high, implying that the tracers were correctly identified.
We have identified and selected a well isolated and distinct 63 XBPs 
which are distributed over the different heliographic latitudes.  Using SSW in IDL, we have generated
the full-disk maps of the images and overlayed them with longitude and latitude grid maps.  
Using the tracer method (the tracer method is based on a visual identification of a particular XBP 
that can be used as a tracer), coronal X-ray bright points are visually traced in consecutive images on a computer
screen.  This was carried out several times.  The tracing of XBPs was performed in 211 consecutive
images for the observing period of January, March and April, 2007.  The rotation velocities were
determined by linear least-squares fit of the central meridian distances (CMD) as a function of time, and more
images were used to determine the velocity of each XBP.  We have measured a large number of rotation
velocities for different possible CMDs in each latitude band, so that more number of data points have
been used in the analysis.  We have compared the 
angular rotational velocity values as a function of latitude of the corresponding XBPs.  

\subsection{Yohkoh/SXT full-disk images}

A multiple observing sequences of the Yohkoh/SXT full-frame images
selected for 3-months in each year for the time interval from 1992 to 2001 have been used for the
analysis.  From the
daily images we have identified the distinct and isolated XBPs and finally selected 134 XBPs.  
The tracer method has been applied to Yohkoh/SXT images (We have already discussed the details of the method
 above for the Hinode/XRT 
images) for all the XBPs as they pass over the solar disk and measured their coordinates.
From these coordinates, the sidereal angular rotation velocities 
and their sizes in terms of arc sec for all the years of data have been calculated.   Finally, we made the 
plots of angular rotation velocity as a function of (i) latitude for all the years and separately for each year 
and (ii) sizes of the XBPs.  Only those XBPs that were present in at least five to six
consecutive images were selected for the rotation velocity and size determination from Hinode/XRT and Yohkoh/SXT 
data sets.   Further details on the method of data analysis are also discussed below together 
with the results and discussions.


\section{Results and discussions}


\subsection{Coronal differential rotation profiles}

 \begin{figure}
   \centering
  \includegraphics[width=13.0cm,height=9.5cm]{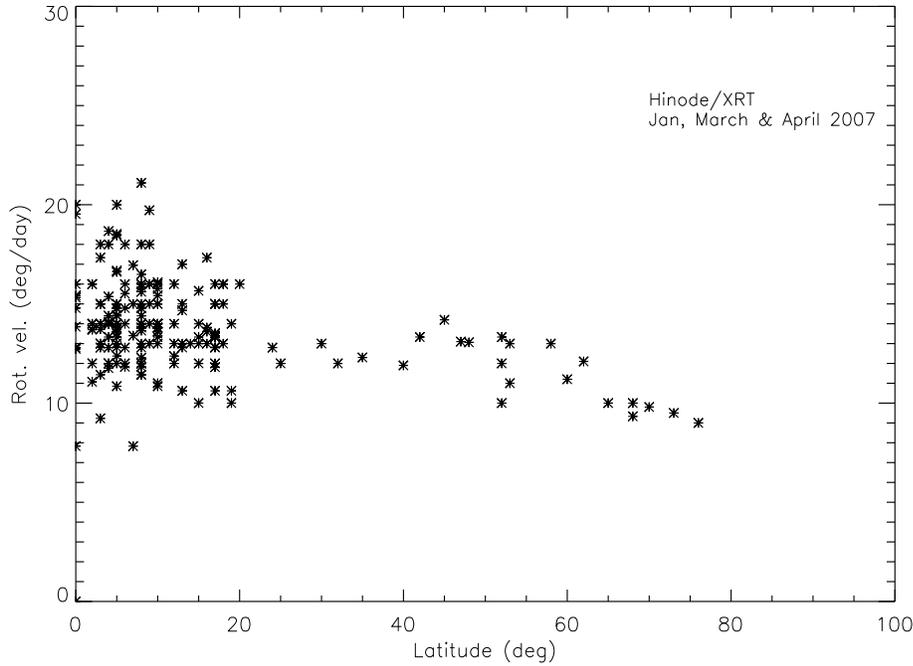}
   \caption{Coronal sidereal angular rotation velocity as a function of latitude to demonstrate that the corona
rotates differentially. The rotation velocity has been measured with the help of XBPs as tracers
from Hinode/XRT full-disk images of January, March, and April 2007.}
\label{FigVibStab}
    \end{figure}

For comparison with previous studies of solar rotation, we fit the data by A + B $sin^2$ $\psi$ +
C $sin^4$ $\psi$, where $\psi$ is the solar latitude and the coefficients A, B, and C listed for Hinode/XRT and
Yohkoh/SXT data in Table 1.  In the additional constraint B=C is assumed to avoid crosstalk
between the parameters B and C.  This is especially important when the data from middle and higher solar
latitudes are considered and when the parameters from different data sets are compared.  However, one way to avoid crosstalk is to obtain the data from all latitudes and in this context XBPs, covering
latitudes from the equator upto about 70 deg, which has an obvious advantage.  The coefficients (from Table 1) 
of the above approximation for the Hinode/XRT data are in general agreement with
previous studies (e.g., Brajsa et al. 2001, 2002 \& 2004) with an equatorial rotation velocity.  The
equatorial rotation velocity of Yohkoh/SXT data is higher than that of Hinode/XRT rotation velocities.
Figure 1 shows a profile of solar rotation of 63 XBPs
derived from the 
images of Hinode/XRT for all the 3-months (January, March \& April, 2007).  Although there is
a significant scatter in the profile, the solar differential rotation is clearly seen.   The scatter in Fig.1
can be attributed to evolution and proper motions of some of the XBPs, as well as error in displacements determination.
 We find less number of data
points at higher latitude in Fig.1 and it is mainly due to absence of XBPs at higher 
latitudinal regions.  Similarly, in Fig.2(a), we have
shown angular rotation velocity for all the 134 XBPs as a function of latitude derived from Yohkoh/SXT images for the 
years from 1992 to 2001.  We find that the latitudinal profile of solar rotation derived from XBPs in
 Hinode/XRT images is similar to the one derived independently from XBPs of Yohkoh/SXT full-disk images and
in both the cases the differential rotation is clearly seen.  In addition, we
infer from Fig.1 and Fig.2(a) that the equatorial sidereal angular rotation velocity in the case of
Yohkoh/SXT (17.6 deg/day from Table 1) 
is higher than in Hinode XBPs (14.2 deg/day), coronal bright points (14.7 deg/day, Brajsa et al. 2004), and in
sunspot regions (14.5 deg/day) at the photospheric level (Howard 1984, Balthasar et al. 1986).  However, the reason 
for the presence
of higher rotational velocity with Yohkoh/SXT data compared to other results is not understood.
\begin{table}
\caption{ Variation of the coefficients A, B and their standard errors EA \& EB in deg/day for different tracers}            
\label{table:1}      
\centering                          
\begin{tabular}{c c c c c}        
\hline\hline                 

Tracer & A $\pm$ EA & -B $\pm$ EB & Data Source & References  \\

\hline                        

XBPs &  14.192 $\pm$ 0.170 & 4.211 $\pm$ 0.775 & Hinode/XRT & present work   \\      
XBPs &  17.597 $\pm$ 0.398 & 4.542 $\pm$ 1.136 & Yohkoh/SXT & present work \\
Coronal bright points &  14.677 $\pm$ 0.033   & 3.100 $\pm$ 0.140 & SOHO/EIT & Brajsa et al (2004)          \\
Sunspots & 14.551 $\pm$ 0.006 & 2.870 $\pm$ 0.060 & Greenwich & Balthasar et al. (1986) \\
Sunspots & 14.370 $\pm$ 0.010 & 2.590 $\pm$ 0.160 & Greenwich & Brajsa et al (2002) \\

\hline                                   


\end{tabular}
\end{table}
  
\subsection{Dependence of coronal differential rotation on sizes of XBPs}

In addition to sidereal angular rotation velocity of XBPs, we measured the sizes of all the 134
XBPs in full-disk images observed from Yohkoh/SXT for the time interval from 1992 to 2001.  In order to 
determine the relationship between the sidereal angular rotation velocity and the sizes of the XBPs, we have plotted
the sidereal angular rotation velocity as a function of latitude in Fig.2(a) and as a function of sizes of the XBPs in Fig.2(b). 
The correlation
coefficient between the sidereal angular rotation velocity and the sizes of XBPs is found to be 0.065. 
It is evident 
from the Fig.2 that the coronal sidereal angular rotation velocity does not depend on the sizes of XBPs.  At the same time,
it is interesting to re-investigate how the sidereal angular rotation velocity varies in long-lived and short-lived XBPs in
comparison with photospheric sunspots and surroundings.  We examined the life span and variations in 
sidereal angular rotation velocity
of some of the XBPs from the collection, we find that the sidereal angular rotation velocity values appears
to be not changing with their life time.  This means that the long and short-lived XBPs 
show almost the same rotational velocity.  In contrast,  Golub 
and Vaiana (1978) have shown from their preliminary analysis that the rotation rate of long-lived 
X-ray features is equal to that of sunspots and on the other hand the short-lived X-ray features rotation 
rate is consistent with that of photospheric gas. 
In the case of sunspot data analysis, Balthasar et al. (1986) have concluded that there is a tendency that larger 
sunspot groups will yield a larger equatorial velocities, but this tendency is much weaker than the
dependence on the type of sunspots.   However, the coronal sidereal angular rotation velocity does not 
depend on the sizes of XBPs and we can not extrapolate the relationship between the photospheric rotational velocity and the sizes
of sunspot groups to coronal level.
 \begin{figure}
   \centering
  \includegraphics[width=13.0cm,height=9.5cm]{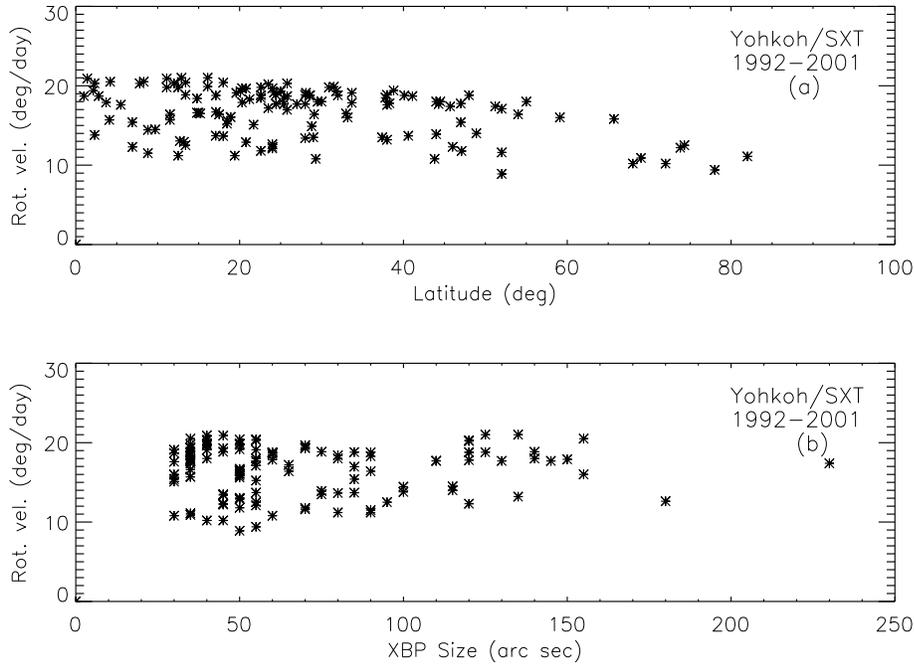}
\caption{Coronal sidereal angular rotation velocity is compared with (a) latitude and (b) sizes of the XBPs derived
from the images of Yohkoh/SXT for the time interval from 1992 to 2001.  The correlation coefficient between the rotation velocity and sizes of
XBPs is 0.065.}
\label{FigVibStab}
    \end{figure}

\subsection{Variation in coronal differential rotation as a function of the phases of the 
solar magnetic cycle}

The coronal X-ray bright points are always present throughout the solar cycle and are more broadly 
distributed over all the latitude regions compared to sunspot regions, so that the differential
rotation can be measured at any phase in the solar cycle.  It is known that the observed XBP number
is anti-correlated with the solar cycle, but this is an observational bias and the number
density of XBPs is nearly independent of the 11-yr solar activity cycle (Nakakubo \& Hara 1999;
Sattarov et al. 2002; Hara \& Nakakubo 2003).
 In Fig.3, we have plotted, for 
Yohkoh/SXT data, the sidereal angular rotation velocity as a function of latitude for
the years from 1992 and 2001 separately.  Each rectangular box in Fig.3 represents the differential 
rotation profile of each year.  It is known that for sunspots at the photospheric level
the mean rotation period varied from cycle to cycle in such a way that the rotation period was
less in low activity cycle and longer rotation period in high activity cycle (Kambry and Nishikawa, 
1990).  On the other hand, Mehta (2005) had analysed radio observational data and found that there is 
no systematic relationship between coronal rotation period and phase of the solar cycle.  We have calculated
the parameters (A and B) of the solar coronal differential rotation for each year separately and 
tabulated them in Table 2.  For
comparison, the yearly mean sunspot numbers are also presented in Table 1.  In
Fig.4, we have plotted the angular rotation velocity against the yearly mean sunspot number and the correlation
coefficient (r) is found to be 0.2.  It is clearly seen from Fig.3, Table 2 and Fig.4 that the
variation in the rotational velocities is independent of the phases of the solar magnetic cycle and the corona rotates differentially in all the years.  In the case 
of photospheric differential rotation using sunspots as tracers, the rotation velocity is anticorrelated with 
the solar activity (as shown in Fig.2 and Table 7 of Balthasar et al. 1986).
\begin{table}
\caption{Variation of the coefficients A \& B and their errors EA \& EB in deg/day for different years representing different phases of the
solar cycle}             
\label{table:1}      
\centering                          
\begin{tabular}{c c c c c}        
\hline\hline                 

Year & A $\pm$ EA & -B $\pm$ EB  & Yearly Mean Sunspot Numbers\\

\hline                        

1992 & 15.296 $\pm$ 1.161  &  0.344 $\pm$ 3.698 & 93.6 & \\

1993 & 13.683 $\pm$ 1.208  &  2.698 $\pm$ 5.203  & 54.5 & \\

1994 &  16.542 $\pm$ 0.985 &   4.261 $\pm$ 2.618 & 31.0 & \\

1995 &  19.547 $\pm$ 0.791  &    8.663 $\pm$ 1.873 & 18.2 & \\

1996 &  15.536 $\pm$ 2.288  &   3.717 $\pm$ 6.665 & 8.4 & \\

1997 &  16.216 $\pm$ 3.139  &   5.638 $\pm$ 6.754 & 20.3 & \\

1998 &  13.552 $\pm$ 2.773  &   -0.175 $\pm$ 8.836 & 62.6 & \\

1999 &   20.359 $\pm$ 1.533  &   12.266 $\pm$ 4.645 & 96.1 & \\

2000 &  19.618 $\pm$ 1.768  &   10.911 $\pm$ 5.486 & 123.3 & \\

2001 &  15.704 $\pm$ 2.062  &   3.984 $\pm$ 7.078 & 123.3 & \\

\hline                                   


\end{tabular}
\end{table}
 \begin{figure}
   \centering
  \includegraphics[width=13.5cm,height=12.5cm]{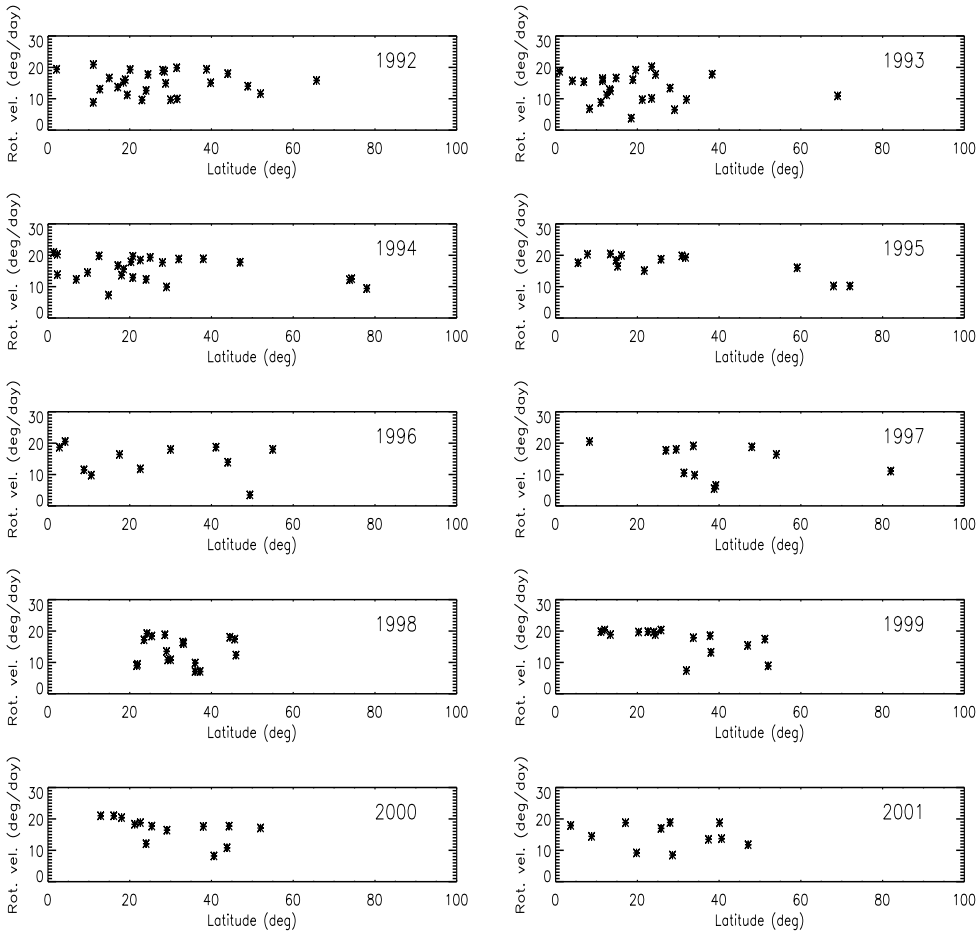}
\caption{Variation of coronal rotation velocity as a function of the phases of the solar
magnetic cycle.  The sidereal angular rotation velocities have been derived from the XBPs in Yohkoh/SXT
images for time interval from 1992 to 2001.}
\label{FigVibStab}
    \end{figure}

  From a detailed analysis of daily full-disk X-ray images obtained using the Hinode/XRT and Yohkoh/SXT experiments,
we conclude that the coronal X-ray bright points are good tracers to determine the angular rotation velocity
of the corona. It is found that the corona rotates differentially like in the photosphere and 
chromosphere.   The coronal sidereal angular rotation velocity does not depend on the sizes of XBPs and the 
differential rotation is present throughout the solar cycle.  The results do not reveal any strong evidence 
for dependence of angular rotation
velocity on solar magnetic cycle.
 \begin{figure}
   \centering
  \includegraphics[width=13.5cm,height=7.0cm]{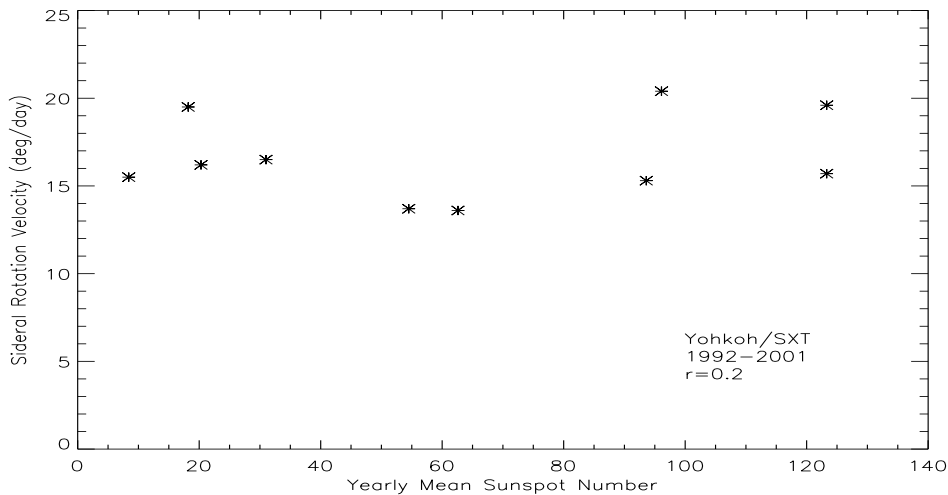}
\caption{Coronal sidereal angular rotation velocity (derived from XBPs of Yohkoh/SXT full-disk images) is 
compared with Yearly Mean Sunspot Number for the years from 1992 to 2001.  The correlation coefficient (r) = 0.2.}
\label{FigVibStab}
    \end{figure}
\begin{acknowledgements}
Hinode is a Japanese mission developed and launched by ISAS/JAXA,
collaborating with NAOJ as
a domestic partner, NASA and STFC (UK) as international partners. Scientific operation of the
Hinode mission is conducted by the Hinode science team organized at ISAS/JAXA. This team mainly
consists of scientists from institutes in the partner countries. Support for the post-launch
operation is provided by JAXA and NAOJ (Japan), STFC (U.K.), NASA (U.S.A.), ESA, and NSC (Norway).
The SXT programme was supported at Lockheed under NASA
contract NAS8-37334 with the Marshall Space Flight Center and by the Lockheed Independent 
Research Programme.  

A part of the Yohkoh/SXT data analysis was done at NAOJ \& ISAS when the author
held a NAOJ Visiting Professorship in 2002.  The author would like to express his sincere thanks to
Tetsuya Watanabe and Takashi Sakurai for the discussion, their kind hospitality and support provided 
during his stay at NAOJ. Many thanks are to Katharine
Reeves and Y. Katsukawa for their help in accessing the Hinode/XRT and Yohkoh/SXT images.
The author is grateful to Ed DeLuca and L. Golub for their valuable suggestions and discussion on the
XRT observations and results.   Thanks to B. A. Varghese for his help at many stages during the analysis 
of X-ray full-disk images and to A. Satyanarayanan for reading through the manuscript and given
valuable suggestions.  The author is thankful to the unknown referee for many constructive
comments and suggestions on the paper.

\end{acknowledgements}

\end{document}